\documentclass[12pt]{article}
\usepackage{amsfonts}
\usepackage{amsmath}
\usepackage{amssymb}
\usepackage{mathptmx}
\usepackage{geometry}
\usepackage{graphicx}
\usepackage{hyperref}
\usepackage{cancel}
\usepackage{textcomp}
\usepackage{tikz}
\usepackage{bm}
  \usepackage{mathrsfs}
  \usepackage{filecontents}
\usepackage{times}
\usepackage{epsfig}
\usepackage{xcolor}
\usepackage{slashed}
\usepackage{booktabs}% build a table
\usepackage{latexsym}
\usepackage{verbatim}
\usepackage{extarrows}
\usepackage{multirow}
\usepackage{rotating}
\usepackage{colortbl}
\usepackage{indentfirst}
\usepackage{float}
\definecolor{mygray}{gray}{.9}
\definecolor{intnull}{RGB}{213,229,255}
\usepackage{arydshln}
\usepackage{diagbox}
\usepackage{makecell}
\usepackage{array}
\usepackage{cite}
\usepackage{caption}
\usepackage{tikz}
\usetikzlibrary{arrows,shapes,chains}
\usepackage{graphicx, subfig}
\begin{document}
\renewcommand{\thefootnote}{\fnsymbol{footnote}}
\baselineskip=16pt
\pagenumbering{arabic}
\vspace{1.0cm}
\begin{center}
{\Large\sf Connections between entropy surface density and microscopic property in black holes}
\\[10pt]
\vspace{.5 cm}

{Xin-Chang Cai\footnote{E-mail address: caixc@mail.nankai.edu.cn} and Yan-Gang Miao\footnote{Corresponding author. E-mail address: miaoyg@nankai.edu.cn}}
\vspace{3mm}

{School of Physics, Nankai University, Tianjin 300071, China}

\vspace{4.0ex}

{\bf Abstract}
\end{center}

We introduce a new microscopic quantity $\varepsilon $ that describes the contribution of a single black hole molecule to black hole entropy and give a key relation that this microscopic quantity is proportional to the macroscopic quantity --- entropy surface density $\sigma_{S}$, thus connecting a microscopic quantity to a macroscopic quantity of black holes. Such a connection provides a new probe for understanding the black hole microstructure from the macroscopic perspective. We also classify black holes in terms of entropy surface density. When its entropy surface density is larger (smaller) than $1/4$, this type of black holes is defined as strong (weak) Bekenstein-Hawking black holes, where the black holes with $1/4$-entropy surface density are regarded as Bekenstein-Hawking black holes. We compute the entropy surface density for four models, where three of them are regular black holes --- the Bardeen black hole, the Ay\'{o}n-Beato-Garc\'{\i}a black hole, and the Hayward black hole, and one of them is a singular black hole --- the five-dimensional neutral Gauss-Bonnet black hole. Under this scheme of classification,
the Bardeen black hole, the Ay\'{o}n-Beato-Garc\'{\i}a black hole, and the five-dimensional neutral Gauss-Bonnet black hole belong to the type of strong Bekenstein-Hawking black holes, but the Hayward black hole belongs to the type of weak Bekenstein-Hawking black holes, and the four black holes approach to the Bekenstein-Hawking  black hole if their event horizon radii go to infinity. In addition, we find interesting properties in phase transitions from the viewpoint of entropy surface density, or equivalently from the viewpoint of a single black hole molecule entropy, for the five-dimensional neutral Gauss-Bonnet AdS black hole.
%For the five-dimensional neutral Gauss-Bonnet AdS black hole, we notice that its entropy surface density changes suddenly when passing through the small-large coexistence curve, which means that there is a big change in the contribution of a single molecule's entropy to the whole entropy. Moreover, we note that as the temperature increases, the contribution of a large black hole molecule to the entropy increases slowly,  while the contribution of a small black hole molecule to the entropy  decreases significantly, and the contribution of small black hole molecules to the entropy is always greater than that of large black hole molecules until they have the same contribution at the critical point.

\newpage

\section{Introduction}
Black holes are a kind of magical celestial bodies predicted at first by Einstein's general relativity. Their essence is a spacetime region that even light cannot escape. Recently, the gravitational waves detected by LIGO and Virgo collaborations~\cite{P1,P2} and the first image of the supermassive black hole in the center of galaxy M87 detected by the Event Horizon Telescope (EHT) collaboration~\cite{P3,P4} have opened a new window for black hole physics.

Black hole thermodynamics has been regarded as an important branch of black hole physics since Hawking, Bekenstein, and others' pioneering work~\cite{P5,P6,P7,P8}. Black holes not only have~\cite{P9} temperature, entropy, pressure in de Sitter or anti de Sitter spacetime, but also have~\cite{P10,P11} van der Waals-like phase transition in anti de Sitter spacetime if they have electrical charges or magnetic charges. In particular, Ruppeiner thermodynamic geometry~\cite{P12,P13,P14,P15} has been used to study black hole microstructures and fruitful achievements have been obtained~\cite{P16,P17,P18,P19,P20,P21,P22,P23}. The results show~\cite{P13,P15,P19,P21} that the Ruppeiner thermodynamic curvature scalar corresponds to repulsive (attractive) or no interactions among black hole molecules when it is larger (smaller) than or equal to zero. Especially, the number density~\cite{P19} related to event horizons can measure the microscopic degrees of freedom of black holes and suffers a sudden change in the small-large black hole phase transition. Obviously, black holes as a thermodynamic system have rich microstructures.

As is well known, many black holes in Einstein's gravity satisfy the famous Bekenstein-Hawking entropy relation, that is, $S={A}/{4}$, where $A$ is event horizon area. However, it is not valid~\cite{P33,P34,P35,P36,P37,P24,P82,P83,P25,P26,P28,P29,P30,P31,P32}, for instance, in regular black holes~\cite{P38,P39,P80} and Gauss-Bonnet black holes~\cite{P36,P37,PX10,PDT1}, where the former is a kind of black holes without spacetime singularity and the latter a special kind of  black holes in Lovelock gravity theory with higher-order curvature corrections. This phenomenon indicates that the entropy surface density, $\sigma_{S} \equiv {S}/{A}$, of regular black holes and Gauss-Bonnet black holes is not a constant ${1}/{4}$, but a function of  event horizons, and that the entropy surface density of regular and Gauss-Bonnet black holes may be a non-trivial physical quantity.

In this paper, we take three regular black holes and one singular black hole as examples to study the relationship between the macroscopic quantity --- entropy surface density and microscopic properties of black holes, such as a single black hole molecule entropy and thermodynamic phase transitions. The three regular black holes are the Bardeen black hole, the Ay\'{o}n-Beato-Garc\'{\i}a black hole, and the Hayward black hole, and the singular black hole is the five-dimensional neutral Gauss-Bonnet black hole. Our main motivations are as follows.

(i) From the perspective of microstructure, black holes can be regarded as a thermodynamic system composed of black hole molecules, so it is necessary to explore the influence of black hole molecules on entropy and compare contributions from different black hole molecules.

(ii) It is important to know whether the entropy surface density is a non-trivial physical quantity and how it is  associate with thermodynamic phase transitions.

The paper is organized as follows. In Sec. 2, we give a general relation about the entropy surface density and the contribution of a single black hole molecule to entropy, and calculate the entropy surface density for the four black holes. Then, we discuss in Sec. 3 thermodynamic phase transitions from the perspective of entropy surface density in the five-dimensional neutral Gauss-Bonnet AdS black hole.  Finally, we make a simple summary in Sec. 4. We use the units $c=G=k_{B}=\hbar=1$ and the sign convention $(-,+,+,+)$ throughout this paper.

\section{Relationship between entropy surface density and a single black holes molecule entropy}
The entropy surface density of black holes is defined as
\begin{equation}
\label{1}
\sigma _{S}\equiv \frac{S}{A},
\end{equation}
where $S$ is entropy and $A$ event horizon area of black holes. Considering that every  microscopic degree of freedom of black holes is carried by Planck area pixels~\cite{P13,P19} and assuming that each black hole molecule has the same microscopic degrees of freedom, we deduce that the total number of black hole molecules $N_{B}$ is proportional to the total number of black hole microscopic degrees of freedom $N$,
\begin{equation}
\label{3}
N_{B}=kN=\frac{kA}{ l_{P}^{D-2}},
\end{equation}
where $l_{P}$ is Planck length, $D$ spacetime dimension, and $k$ a positive constant coefficient. We define a microscopic quantity $\varepsilon $, which is used to describe the contribution of a single black hole molecule to the  black hole entropy $S$,
\begin{equation}
\label{4}
\varepsilon \equiv \frac{S}{N_{B}}.
\end{equation}
Combining Eqs.~(\ref{1}), (\ref{3}), and (\ref{4}), we obtain
\begin{equation}
\label{5}
\varepsilon \equiv \frac{S}{N_{B}}=\frac{S/A}{N_{B}/A}=\frac{\sigma_{S} }{\frac{k}{ l_{P}^{D-2}}}~\ ~\ ~\ \Longrightarrow ~\ ~\ ~\
\sigma_{S}=\frac{k}{ l_{P}^{D-2}}\varepsilon .
\end{equation}

From Eq.~(\ref{5}), we deduce an important relationship between the macroscopic quantity --- entropy surface density $\sigma_{S}$ and the microscopic quantity $\varepsilon $, that is, $\sigma_{S}$ is proportional to $\varepsilon $.
The entropy surface density can measure the contribution of a single black hole molecule to the black hole entropy, thus revealing an intrinsic connection between the entropy surface density and the black hole microstructure. This result provides a new probe for us to understand the black hole microstructure from the macroscopic perspective.

\subsection{Entropy surface density of regular black holes}

A static spherically symmetric black hole can be described by the line element,
\begin{equation}
\label{7}
ds^{2}=-f_{i}(r)dt^{2}+\frac{dr^{2}}{f_{i}(r)}+r^{2}\left(d\theta^{2}+\sin^{2}\theta d\varphi^{2}\right),  ~\ ~\  ~\ (i=1,2,3)
\end{equation}
where $f_{1}(r)$, $f_{2}(r)$, and $f_{3}(r)$ represent~\cite{P38,P39,P40} the metric functions of the Bardeen black hole, the Ay\'{o}n-Beato-Garc\'{\i}a black hole, and the Hayward black hole, respectively,
\begin{equation}
\label{8}
f_{1}(r)=1-\frac{2Mr^{2}}{(r^{2}+g_{1}^{2})^{{3}/{2}}},
\end{equation}
\begin{equation}
\label{9}
f_{2}(r)=1-\frac{2Mr^{2}}{(r^{2}+q^{2})^{{3}/{2}}}+\frac{q^{2}r^{2}}{(r^{2}+q^{2})^{2}},
\end{equation}
\begin{equation}
\label{10}
f_{3}(r)=1-\frac{2Mr^{2}}{r^{3}+g_{2}^{3}},
\end{equation}
where $M$ is black hole mass, $g_{1}$ magnetic charge,  $q$ electric charge,  and $g_{2}$ integration constant  related to magnetic charges or simply regarded as~\cite{P41} magnetic charge.

The entropy and event horizon area of the three regular black holes take~\cite{P83,P28,P29} the forms,\footnote{According to Ref.~\cite{P29}, the logarithm term of the Bardeen black hole has been rewritten.}  respectively,
\begin{equation}
\label{11}
S_{1}=\pi g_{1}^{2}\left[\left(\frac{r_{\rm H}}{g_{1}}-\frac{2g_{1}}{r_{\rm H}}\right)\sqrt{1+\left(\frac{r_{\rm H}}{g_{1}}\right)^{2}}
+3\ln\left(\frac{r_{\rm H}}{g_{1}}+\sqrt{1+\left(\frac{r_{\rm H}}{g_{1}}\right)^{2}}\right)\right],
\end{equation}
\begin{equation}
\label{12}
S_{2}=\pi q^{2}\left[\left(\frac{r_{\rm H}}{q}-\frac{2q}{r_{\rm H}}\right)\sqrt{1+\left(\frac{r_{\rm H}}{q}\right)^{2}}
+3\ln\left(\frac{r_{\rm H}}{q}+\sqrt{1+\left(\frac{r_{\rm H}}{q}\right)^{2}}\right)\right],
\end{equation}
\begin{equation}
\label{13}
S_{3}=\pi r_{\rm H}^{2}-\frac{2\pi g_{2}^{3}}{r_{\rm H}},
\end{equation}
\begin{equation}
\label{14}
A_{1}=A_{2}=A_{3}=4\pi r_{\rm H}^{2},
\end{equation}
where $r_{\rm H}$ is event horizon radius. Correspondingly, the extreme event horizon radii have~\cite{P39,P42,P43} the forms, respectively,
\begin{equation}
\label{15}
r_{\rm EH1}=\sqrt{2}\,g_{1},
\end{equation}
\begin{equation}
\label{16}
r_{\rm EH2}=\left[\frac{1}{3} \left(10 \sqrt[3]{\frac{2}{3 \sqrt{321}+83}}+\sqrt[3]{\frac{1}{2} \left(3 \sqrt{321}+83\right)}+1\right)\right]^{1/2}q,
\end{equation}
\begin{equation}
\label{17}
r_{\rm EH3}=\sqrt[3]{2}\,g_{2}.
\end{equation}

Combining Eqs.~(\ref{11}-\ref{17}) with Eq.~(\ref{1}), we derive the entropy surface density for the three regular black holes as follows,
\begin{equation}
\label{18}
\sigma _{S_{1}}=\frac{1}{4}\left(\frac{g_{1}}{r_{\rm H}}\right)^{2}\left[\left(\frac{r_{\rm H}}{g_{1}}-\frac{2g_{1}}{r_{\rm H}}\right)
\sqrt{1+\left(\frac{r_{\rm H}}{g_{1}}\right)^{2}}
+3\ln\left(\frac{r_{\rm H}}{g_{1}}+\sqrt{1+\left(\frac{r_{\rm H}}{g_{1}}\right)^{2}}\right)\right],
\end{equation}
\begin{equation}
\label{19}
\sigma _{S_{2}}= \frac{1}{4}\left(\frac{q}{r_{\rm H}}\right)^{2}\left[\left(\frac{r_{\rm H}}{q}-\frac{2q}{r_{\rm H}}\right)
\sqrt{1+\left(\frac{r_{\rm H}}{q}\right)^{2}}+3\ln\left(\frac{r_{\rm H}}{q}
+\sqrt{1+\left(\frac{r_{\rm H}}{q}\right)^{2}}\right)\right],
\end{equation}
\begin{equation}
\label{20}
\sigma _{S_{3}}=\frac{1}{4}-\frac{1}{2}\left(\frac{g_{2}}{r_{\rm H}} \right )^{3}.
\end{equation}

For the Bardeen black hole, we draw in Fig. 1 the curve of the entropy surface density $\sigma _{S_{1}}$ as a function of the reduced event horizon radius ${r_{\rm H}}/{g_{1}}$. We can see that the entropy surface density increases rapidly to the maximum, $\sigma_{S_{1}}^{\rm max}\approx 0.432$, at the event horizon radius $r_{\rm H}\approx 1.509\,g_{1}$ when the reduced event horizon radius increases from its extreme value, $r_{\rm EH1}=\sqrt{2}\,g_{1}$. Then, it decreases and eventually approaches ${1}/{4}$ when ${r_{\rm H}}/{g_{1}}$ continues to increase and goes to infinity. This phenomenon means that the contribution of the Bardeen black hole molecules to entropy increases rapidly to the maximum value at the beginning, then decreases, and finally tends to ${1}/{4}$ with the increase of ${r_{\rm H}}/{g_{1}}$.

\begin{figure}
\centering
\begin{minipage}[t]{0.8\linewidth}
\centering
\includegraphics[width=100mm]{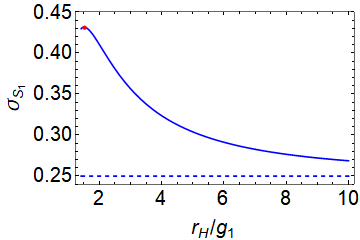}
\caption*{Figure 1. The entropy surface density $\sigma_{S_{1}}$ with respect to the reduced event horizon radius ${r_{\rm H}}/{g_{1}}$ for the Bardeen black hole. Here the red dot corresponds to the maximum entropy surface density, $\sigma_{S_{1}}^{\rm max}\approx 0.432$, at the event horizon radius, $r_{\rm H}\approx 1.509\,g_{1}$. The dotted line represents the entropy surface density of the Schwarzschild black hole as a comparison.}
\label{fig1}
\end{minipage}
\end{figure}

For the Ay\'{o}n-Beato-Garc\'{\i}a black hole, we draw in Fig. 2 the curve of the entropy surface density $\sigma _{S_{2}}$ as a function of the reduced event horizon radius ${r_{\rm H}}/{q}$. We can see that the entropy surface density $\sigma _{S_{2}}$ decreases monotonically from its maximum value, $\sigma_{S_{2}}^{\rm max}\approx 0.431$, at the extreme event horizon radius, $r_{\rm EH2}\approx 1.585q$, and eventually approaches ${1}/{4}$ when ${r_{\rm H}}/{q}$ increases and goes to infinity. This phenomenon indicates that the contribution of the  Ay\'{o}n-Beato-Garc\'{\i}a black hole molecules to entropy decreases monotonically and finally tends to ${1}/{4}$ with the increase of ${r_{\rm H}}/{q}$.

\begin{figure}
\centering
\begin{minipage}[t]{0.8\linewidth}
\centering
\includegraphics[width=100mm]{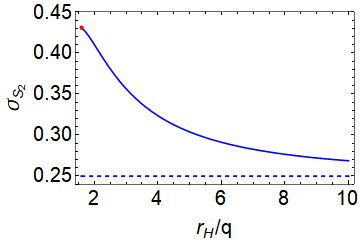}
\caption*{Figure 2. The entropy surface density $\sigma_{S_{2}}$ with respect to the reduced event horizon radius ${r_{\rm H}}/{q}$ for the Ay\'{o}n-Beato-Garc\'{\i}a black hole. Here the red dot corresponds to the maximum entropy surface density, $\sigma_{S_{2}}^{\rm max}\approx 0.431$, at the extreme event horizon radius, $r_{\rm EH2}\approx 1.585q$. The dotted line represents the entropy surface density of the Schwarzschild black hole as a comparison.}
\label{fig1}
\end{minipage}
\end{figure}

For the Hayward black hole, we draw in Fig. 3 the curve of the entropy surface density $\sigma _{S_{3}}$ as a function of the reduced event horizon radius ${r_{\rm H}}/{g_{2}}$. We can see that the entropy surface density $\sigma _{S_{3}}$ increases monotonically from its minimum value, $\sigma_{S_{3}}^{\rm min}=0$, at the extreme event horizon radius, $r_{\rm EH3}=\sqrt[3]{2}\,g_{2}$, and eventually approaches ${1}/{4}$ when ${r_{\rm H}}/{g_{2}}$ increases and goes to infinity. This phenomenon implies that the contribution of the Hayward  black hole molecules to entropy increases monotonically and finally tends to ${1}/{4}$ with the increase of ${r_{\rm H}}/{g_{2}}$.

\begin{figure}
\centering
\begin{minipage}[t]{0.8\linewidth}
\centering
\includegraphics[width=100mm]{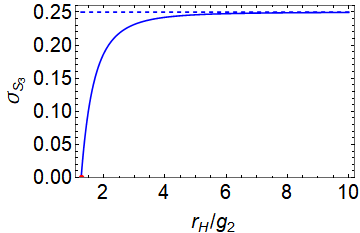}
\caption*{Figure 3. The entropy surface density $\sigma_{S_{3}}$ with respect to the reduced event horizon radius ${r_{\rm H}}/{g_{2}}$ for the Hayward black hole. Here the red dot corresponds to the minimum entropy surface density, $\sigma_{S_{3}}^{\rm min}=0$, at the extreme event horizon radius, $r_{\rm EH3}=\sqrt[3]{2}\,g_{2}$. The dotted line  represents the entropy surface density of the Schwarzschild black hole as a comparison.}
\label{fig1}
\end{minipage}
\end{figure}

\subsection{Entropy surface density of the five-dimensional neutral Gauss-Bonnet black hole}

A static spherically symmetric five-dimensional neutral Gauss-Bonnet black hole can be described~\cite{P36,PX10,PDT1,P51} by the line element,
\begin{equation}
\label{100}
ds^{2}=-f(r)dt^{2}+\frac{dr^{2}}{f(r)}+r^{2}\left(d\theta^{2}+\sin^{2}\theta d\varphi^{2}+\sin^{2}\theta \sin^{2}\varphi d\phi^{2}\right),
\end{equation}
with
\begin{equation}
\label{101}
f(r)=1+\frac{r^{2}}{2\alpha}\left(1-\sqrt{1+\frac{32M\alpha }{3\pi r^{4}}}\right),
\end{equation}
where $\alpha$ is Gauss-Bonnet coupling constant whose dimension is square of length, and $M$ black hole mass. The entropy and event horizon area take~\cite{P36} the forms,
\begin{equation}
\label{102}
S_{4}=\frac{\pi^2}{2}\left(r_{\rm H}^3+6 \alpha r_{\rm H}\right),
\end{equation}
\begin{equation}
\label{103}
A_{4}=2 \pi ^2 r_{\rm H}^3.
\end{equation}
Combining Eqs.~(\ref{102}) and (\ref{103}) with Eq.~(\ref{1}), we obtain the entropy surface density for the five-dimensional neutral Gauss-Bonnet black hole,
\begin{equation}
\label{104}
\sigma_{S_{4}}=\frac{1}{4}+\frac{3}{2} \left(\frac{\sqrt{\alpha}}{r_{\rm H}}\right)^2.
\end{equation}

For the five-dimensional neutral Gauss-Bonnet black hole, we draw in Fig. 4 the curve of the entropy surface density $\sigma_{S_{4}}$ as a function of the reduced event horizon radius ${r_{\rm H}}/{\sqrt{\alpha}}$. We can see that the entropy surface density $\sigma_{S_{4}}$ decreases monotonically and eventually approaches ${1}/{4}$ when  ${r_{\rm H}}/{\sqrt{\alpha}}$ increases and goes to infinity. This phenomenon indicates that the contribution of the black hole molecules to entropy decreases monotonically and finally tends to ${1}/{4}$ with the increase of ${r_{\rm H}}/{\sqrt{\alpha}}$. We note that the entropy surface density will diverge due to the existence of  spacetime singularity when ${r_{\rm H}}/{\sqrt{\alpha}}$ tends to zero (the extreme horizon radius), which is completely different from the case in regular black holes.

\begin{figure}
\centering
\begin{minipage}[t]{0.8\linewidth}
\centering
\includegraphics[width=100mm]{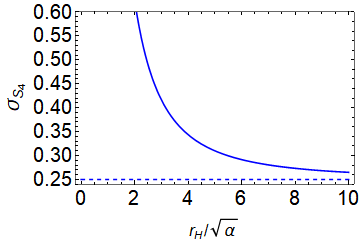}
\caption*{Figure 4. The entropy surface density $\sigma _{S_{4}}$ with respect to the reduced event horizon radius   ${r_{\rm H}}/{\sqrt{\alpha}}$ for the five-dimensional neutral Gauss-Bonnet black hole. The dotted line  represents the entropy surface density of the Schwarzschild black hole as a comparison.}
\label{fig1}
\end{minipage}
\end{figure}

If the black holes with ${1}/{4}$-entropy surface density are called Bekenstein-Hawking black holes, we classify black holes according to the entropy surface density $\sigma_{S}$ as follows:
\begin{itemize}
\item  When $\sigma _{S}>\frac{1}{4}$, the  black holes are called strong Bekenstein-Hawking  black holes.
%\item  When $\sigma _{S}=\frac{1}{4}$, the  black holes are called  Bekenstein-Hawking  black holes.
\item   When $\sigma _{S}<\frac{1}{4}$, the black holes are called weak Bekenstein-Hawking  black holes.
\end{itemize}
If it is assumed that different black holes in the same spacetime dimension have the same coefficient $k$, see Eq.~(\ref{3}), we can alternatively classify black hole molecules in terms of Eq.~(\ref{5}), that is, a strong (weak) Bekenstein-Hawking black hole is composed of strong (weak) Bekenstein-Hawking black hole molecules, where a Bekenstein-Hawking black hole is composed of Bekenstein-Hawking  black hole molecules.
%\begin{itemize}
%\item  When $\sigma _{S}>\frac{1}{4}$, black hole molecules are called strong Bekenstein-Hawking  black hole molecules.
%\item  When $\sigma _{S}=\frac{1}{4}$, black hole molecules are called  Bekenstein-Hawking  black hole molecules.
%\item   When $\sigma _{S}<\frac{1}{4}$, black hole molecules are called weak Bekenstein-Hawking  black hole molecules.
%\end{itemize}

As a result, we utilize the entropy surface density to classify black holes at the macroscopic level, and simultaneously use a single black hole molecule entropy to classify black hole molecules at the microscopic level. In terms of this classification, we know that the Bardeen black hole, the  Ay\'{o}n-Beato-Garc\'{\i}a black hole, and the five-dimensional neutral Gauss-Bonnet black hole are strong Bekenstein-Hawking black holes composed of  strong Bekenstein-Hawking black hole molecules, while the Hayward black hole is weak Bekenstein-Hawking black hole composed of weak Bekenstein-Hawking black hole molecules. In addition, the three regular black holes and one singular black hole tend towards the Bekenstein-Hawking black hole and their black hole molecules towards the  Bekenstein-Hawking black hole molecules as their event horizon radii go to infinity.

\section{Entropy surface density and thermodynamic phase transition of the five-dimensional neutral Gauss-Bonnet AdS black hole}

A static spherically symmetric five-dimensional neutral Gauss-Bonnet  AdS  black hole can be described~\cite{P36,P37,PX10,PDT1,P51} by the line element,
\begin{equation}
\label{105}
ds^{2}=-f(r)dt^{2}+\frac{dr^{2}}{f(r)}+r^{2}\left(d\theta^{2}+\sin^{2}\theta d\varphi^{2}+\sin^{2}\theta \sin^{2}\varphi d\phi^{2}\right),
\end{equation}
with
\begin{equation}
\label{106}
f(r)=1+\frac{r^{2}}{2\alpha}\left(1-\sqrt{1+\frac{32M\alpha}{3\pi r^{4}}-\frac{16\pi \alpha P}{3}}\right),
\end{equation}
where $\alpha$ is Gauss-Bonnet coupling constant whose dimension is the square of length, $M$ black hole mass, and the pressure term, $P=-\frac{\Lambda}{8\pi}$, from the the cosmological constant $\Lambda$.  In addition, the event horizon area, the entropy, the volume, and the Hawking temperature take the forms~\cite{P36,P37,P23}, respectively,
\begin{equation}
\label{107}
A=2 \pi ^2 r_{\rm H}^3,
\end{equation}
\begin{equation}
\label{108}
S=\frac{\pi^2}{2}\left(r_{\rm H}^3+6 \alpha  r_{\rm H}\right),
\end{equation}
\begin{equation}
\label{109}
V=\frac{\pi ^{2}r_{{\rm H}}^{4}}{2},
\end{equation}
\begin{equation}
\label{110}
 T=\frac{8 \pi  P r_{\rm H}^3+3 r_{\rm H}}{12 \pi  \alpha +6 \pi  r_{\rm H}^2}.
\end{equation}
In addition, the critical values of occurrence of a second-order phase transition for the pressure, the horizon radius, the volume, and the Hawking temperature can be obtained~\cite{P23,P44} as follows,
\begin{equation}
\label{111}
P_{\rm C}=\frac{1}{48\pi \alpha},   \qquad r_{\rm CH}=\sqrt{6\alpha}, \qquad V_{\rm C}=18\pi ^{2}\alpha^{2}, \qquad  T_{\rm C}=\frac{1}{2\pi\sqrt{6\alpha}}.
\end{equation}

By introducing the reduced quantities~\cite{P23,P44}for the pressure, the horizon radius, the volume, and the Hawking temperature,
\begin{equation}
\label{115}
\tilde{P}\equiv \frac{P}{P_{\rm C}}, \qquad \tilde{r}\equiv \frac{r_{\rm H}}{r_{\rm CH}},\qquad \tilde{V}\equiv\frac{V}{V_{\rm C}},\qquad \tilde{T}\equiv\frac{T}{T_{\rm C}},
\end{equation}
one can rewrite~\cite{P23,P44} the equation of state,
\begin{equation}
\label{116}
\tilde{P}=\frac{3 \tilde{r}^2 \tilde{T}-3 \tilde{r}+\tilde{T}}{\tilde{r}^3},
\end{equation}
and the equation of coexistence curve of small and large black hole phases on the $\tilde{P}-\tilde{T}$ plane,
\begin{equation}
\begin{aligned}
\label{117}
\tilde{P}=\frac{1}{2} \left(3-\sqrt{9-8 \tilde{T}^2}\right).
\end{aligned}
\end{equation}
Moreover, one can compute~\cite{P50} the  horizon  radii of the coexisting  small and large black holes,
\begin{equation}
\begin{aligned}
\label{118}
\tilde{r}_{s}=\frac{2 \tilde{T}- \sqrt{12 \tilde{T}^2-6\left(3-\sqrt{9-8 \tilde{T}^2}\right)}}{3-\sqrt{9-8 \tilde{T}^2}},
\end{aligned}
\end{equation}
\begin{equation}
\begin{aligned}
\label{119}
\tilde{r}_{l}=\frac{2 \tilde{T}+\sqrt{12 \tilde{T}^2-6\left(3-\sqrt{9-8 \tilde{T}^2}\right)}}{3-\sqrt{9-8 \tilde{T}^2}}.
\end{aligned}
\end{equation}

Combining Eqs.~(\ref{1}), (\ref{107}), (\ref{108}) and (\ref{111}), we can derive the entropy surface density $\sigma $ and the  critical value $\sigma _{\rm C}$ of occurrence of a phase transition for the five-dimensional neutral Gauss-Bonnet AdS black hole as follows,
\begin{equation}
\label{120}
\sigma= \frac{1}{4}+\frac{3 \alpha }{2 r_{\rm H}^2},   \qquad   \sigma _{\rm C}=\frac{1}{2},
\end{equation}
and define the corresponding reduced entropy surface density $\tilde{\sigma}$ as
\begin{equation}
\label{121}
\tilde{\sigma}\equiv \frac{\sigma}{\sigma _{\rm C}}=\frac{1}{2}\left(1+\frac{1}{\tilde{r}^{2}} \right ).
\end{equation}
Again using Eqs.~(\ref{116}) and (\ref{121}), we obtain the reduced temperature,
\begin{equation}
\label{122}
\tilde{T}=\frac{6 \tilde{\sigma }-3+\tilde{P}}{2 \left(\tilde{\sigma}+1\right) \sqrt{2 \tilde{\sigma}-1}}.
\end{equation}

In Fig. 5, we draw the curve of the reduced temperature $\tilde{T}$ as a function of the reduced entropy surface density $\tilde{\sigma}$ for different values of the reduced pressure $\tilde{P}$ and clearly see that there exists a van der Waals-like phase transition  for $\tilde{P}<1$  on the $\tilde{\sigma} - \tilde{T}$ plane.

\begin{figure}
\centering
\begin{minipage}[t]{0.8\linewidth}
\centering
\includegraphics[width=100mm]{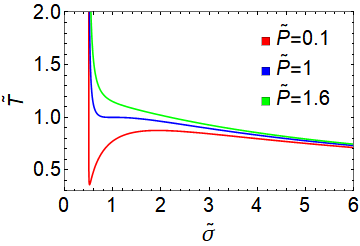}
\caption*{Figure 5. The reduced temperature $\tilde{T}$ with respect to the reduced entropy surface density $\tilde{\sigma}$ for different values of the reduced pressure $\tilde{P}$.}
\label{fig1}
\end{minipage}
\end{figure}

Now combining Eqs.~(\ref{118}), (\ref{119}) and (\ref{121}), we can compute the reduced entropy surface densities on the coexistence curve of small and large phases for the five-dimensional neutral Gauss-Bonnet AdS black hole,
\begin{equation}
\label{123}
\tilde{\sigma}_{s}=\frac{1}{2} \left[1+\frac{\left(3-\sqrt{9-8 \tilde{T}^2}\right)^2}{\left(2 \tilde{T}- \sqrt{12 \tilde{T}^2-6\left(3-\sqrt{9-8 \tilde{T}^2}\right)}\right)^2}\right],
\end{equation}
\begin{equation}
\label{124}
\tilde{\sigma}_{l}=\frac{1}{2} \left[1+\frac{\left(3-\sqrt{9-8 \tilde{T}^2}\right)^2}{\left(2 \tilde{T}+ \sqrt{12 \tilde{T}^2-6\left(3-\sqrt{9-8 \tilde{T}^2}\right)}\right)^2}\right].
\end{equation}

In Fig. 6, we draw the curve of the reduced entropy surface densities, $\tilde{\sigma}_{s}$ and $\tilde{\sigma}_{l}$, with respect to the reduced temperature $\tilde{T}$ on the coexistence curve of small and large black hole phases. We can see that $\tilde{\sigma}_{l}$ slowly increases, while $\tilde{\sigma}_{s}$  decreases significantly, when $\tilde{T}$ increases. This implies that the contribution of a single large black hole molecule to entropy increases slowly, while the contribution of a single small black hole molecule to entropy decreases significantly with the increasing of temperature. When the reduced temperature reaches the critical point, $\tilde T_{\rm C}=1$ together with $\tilde P_{\rm C}=1$, see Eqs.~(\ref{111}), (\ref{115}), and (\ref{117}), i.e., at the end of the coexistence curve of small and large phases (on the $\tilde{P}-\tilde{T}$ plane of Eq.~(\ref{117})), $\tilde{\sigma}_{s}$ and $\tilde{\sigma}_{l}$ equal, that is, they have the same contribution to entropy.

\begin{figure}
\centering
\begin{minipage}[t]{0.8\linewidth}
\centering
\includegraphics[width=100mm]{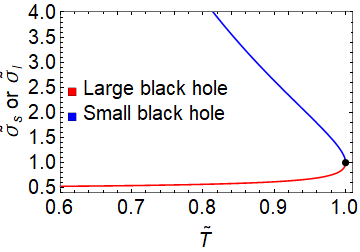}
\caption*{Figure 6. The reduced entropy surface densities, $\tilde{\sigma}_{s}$ and $\tilde{\sigma}_{l}$, with respect to the reduced temperature $\tilde{T}$ on the coexistence curve of small and large black hole phases. Here the black dot represents the critical point with $\tilde T_{\rm C}=1$ and $\tilde{\sigma}_{s}=\tilde{\sigma}_{l}=1$.}
\label{fig1}
\end{minipage}
\end{figure}

Alternatively, the properties of entropy surface densities just mentioned above can be shown clearly in the difference $\Delta \tilde{\sigma }$ between $\tilde{\sigma}_{s}$ and $\tilde{\sigma}_{l}$. In terms of Eqs.~(\ref{123}) and (\ref{124}), we calculate this difference,
\begin{equation}
\label{125}
\Delta \tilde{\sigma }=\tilde{\sigma}_{s}-\tilde{\sigma}_{l}=\sqrt{\frac{3}{2}}\tilde{T} \,\frac{\left(3-\sqrt{9-8 \tilde{T}^2}\right)^2 \sqrt{2 \tilde{T}^2-\left(3-\sqrt{9-8 \tilde{T}^2}\right)}}{8 \tilde{T}^4-12 \left(6-\sqrt{9-8 \tilde{T}^2}\right) \tilde{T}^2+27 \left(3-\sqrt{9-8 \tilde{T}^2}\right)}.
\end{equation}

In Fig. 7, we draw the curve of the difference $\Delta \tilde{\sigma }$ of entropy surface density with respect to  the reduced temperature $\tilde{T}$ on the coexistence curve of the small and large phases for the five-dimensional neutral Gauss-Bonnet AdS black hole. We can see that $\Delta \tilde{\sigma }$ is non-negative and monotonically decreases to zero at the critical point when the reduced temperature increases and approaches $\tilde T_{\rm C}=1$. This phenomenon shows
%that the contribution of black hole molecules to entropy decreases suddenly with the increasing of the reduced temperature. On the other hand, it means
that the contribution of a single small black hole molecule to entropy is always greater than that of a single large black hole molecule, and that one small black hole molecule finally has the same contribution as that of one large black hole molecule at the critical point.
Alternatively, when this AdS black hole system crosses the coexistence curve on the $\tilde{P}-\tilde{T}$ plane at a fixed temperature, the small black hole phase becomes the large black hole phase suddenly, which gives rise to a sudden change in the contribution of the black hole molecules to entropy due to $\Delta \tilde{\sigma }\ne 0$.

\begin{figure}
\centering
\begin{minipage}[t]{0.8\linewidth}
\centering
\includegraphics[width=100mm]{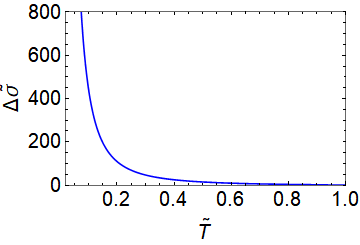}
\caption*{Figure 7. The difference $\Delta \tilde{\sigma}$ between $\tilde{\sigma}_{s}$ and $\tilde{\sigma}_{l}$ with respect to the reduced temperature $\tilde{T}$ on the coexistence curve of small and large phases for the five-dimensional neutral Gauss-Bonnet AdS black hole.}
\label{fig1}
\end{minipage}
\end{figure}

\section{Conclusion }

In this paper, we introduce a new microscopic quantity $\varepsilon $ to describe the contribution of a single black hole molecule to black hole entropy, and establish a key equation that this microscopic quantity is proportional to the macroscopic quantity --- entropy surface density. Thus, we reveal the intrinsic connection between the entropy surface density and the black hole microstructure, which provides a new probe for understanding the black hole microstructure from the macroscopic perspective. Moreover, we propose a scheme to classify black holes by regarding the entropy surface density as a characteristic parameter. If its entropy surface density is larger (smaller) than $1/4$, this type of black holes is called the strong (weak) Bekenstein-Hawking black holes, where the black holes with $1/4$-entropy surface density are defined as Bekenstein-Hawking black holes. Based on this scheme, the Bardeen black hole, the Ay\'{o}n-Beato-Garc\'{\i}a black hole, and the five-dimensional neutral Gauss-Bonnet black hole belong to strong Bekenstein-Hawking black holes composed of strong Bekenstein-Hawking black hole molecules, while the Hayward black hole belongs to a weak Bekenstein-Hawking black hole composed of weak Bekenstein-Hawking black hole molecules.
Quite interesting is that the strong and weak Bekenstein-Hawking black holes tend towards the Bekenstein-Hawking black hole when their event horizon radii go to infinity.

In addition, we analyze the van der Waals-like phase transition for the five-dimensional neutral Gauss-Bonnet AdS black hole from the point of view of entropy surface density, or equivalently, from the point of view of a single black hole molecule entropy.  When this black hole system evolves along the coexistence curve of small and large black hole phases on the $\tilde{P}-\tilde{T}$ plane with an increasing $\tilde{T}$, the entropy of a single small black hole molecule is always greater than that of a single large black hole molecule, and the former equals the latter at the critical point.
Moreover, when the system crosses this coexistence curve, there exists a sudden change in the contribution of the Gauss-Bonnet AdS black hole molecules to entropy.

%We also study the relationship between the entropy surface density of  and the thermodynamic phase transition. We find that as the reduced temperature increases, the difference between the  reduced  entropy surface density of the small and large  black hole along  the small and large five-dimensional neutral Gauss-Bonnet AdS black hole  coexistence curve is non-negative and monotonically decreases until the critical point is zero, which not only shows that there is a sudden change in the contribution of   Gauss-Bonnet AdS  black hole molecules to the entropy  when passing through the  coexistence curve, but also means that  as the temperature increases, the difference between the contribution  of the small and large Gauss-Bonnet AdS black hole molecules to the entropy decreases, and the contribution of  small black hole molecules to the entropy is always greater than that of large black hole molecules until they have the same contribution at the critical point. In addition, we also see that as the reduced temperature $\tilde{T}$ increases, the reduced entropy surface density $\tilde{\sigma}$ of large black holes increases slowly, while that of small black holes decreases significantly, which also means that with the increase of the temperature, the contribution of a large black hole molecule to the entropy increases slowly, while the contribution of a small black hole molecule to the entropy decreases significantly.

In a word, the entropy surface density is a non-trivial physical quantity related to microscopic properties of black holes, such as a single black hole molecule entropy and a phase transition. Based on this physical quantity, we can deepen our understanding of black holes. To this end, it is worth investigating the entropy surface density of rotating black holes, which will be reported in our future work.

\section*{Acknowledgments}

This work was supported in part by the National Natural Science Foundation of China under Grant No. 11675081.

%The authors would like to thank the anonymous referee for the helpful comments that improve this work greatly.

%\newpage

\end{document}